# Demonstration of Robust Chiral Edge Transport in Field-Induced Chern Insulator $MnBi_2Te_4$ Devices with Engineered Geometric Defects

Pinyuan Wang[1,2], Jun Ge[1,2,*], Jiawei Luo[3], Xiaoqi Liu[1,2], Jiayi Yang[1,2], Fucong Fei[4], Fengqi Song[5,6], and Jian Wang[1,2,7,8,†]

[1]*International Center for Quantum Materials, School of Physics, Peking University, Beijing 100871, China*
[2]*Beijing Key Laboratory of Quantum Devices, Peking University, Beijing 100871, China*
[3]*State Key Laboratory of Quantum Functional Materials, and ShanghaiTech Laboratory for Topological Physics, School of Physical Science and Technology, ShanghaiTech University, Shanghai 201210, China*
[4]*National Laboratory of Solid State Microstructures, Collaborative Innovation Center of Advanced Microstructures, and School of Advanced Manufacturing Engineering, Nanjing University, Suzhou 215163, China*
[5]*National Laboratory of Solid State Microstructures, Collaborative Innovation Center of Advanced Microstructures, and School of Physics, Nanjing University, Nanjing 210093, China*
[6]*Atom Manufacturing Institute, Nanjing 211806, China*
[7]*Hefei National Laboratory, Hefei 230088, China*
[8]*Collaborative Innovation Center of Quantum Matter, Beijing 100871, China*

Abstract: Chiral edge states in Chern insulators are theoretically predicted to propagate unidirectionally along the sample boundary with inherent robustness against local perturbations, which manifests as the immunity to impurity-induced backscattering, a key factor for the development of robust, high-performance quantum devices. However, the direct experimental verification of the robustness of chiral edge states remains scarce. Here, we experimentally validate the robustness of the chiral edge states in $MnBi_2Te_4$ devices by introducing engineered geometric defects via atomic force microscopy (AFM) nanomachining. Specifically, under a moderate perpendicular magnetic field, the $MnBi_2Te_4$ devices exhibit the Chern insulator state, characterized by a quantized Hall plateau and simultaneously vanishing longitudinal resistance. To verify the robustness of this topological state, we modify the device geometry by creating a structural cut that reaches the substrate using AFM nanomachining. Remarkably, the quantization behavior survives this drastic modification. The robust nature of the chiral edge transport is further confirmed by two-terminal, three-terminal, and nonlocal measurements, demonstrating that the edge currents can survive the engineered cut. Our results provide strong evidence for the robustness of chiral edge states against geometric disruption and establish AFM nanomachining as a promising technique for topological quantum devices engineering.

Chern insulators (quantum anomalous Hall insulators) possess dissipationless, unidirectional edge transport channels that are prohibited from backscattering by impurities or defects [1–5]. Consequently, this chiral transport is expected to remain robust even against geometric deformation and defects. Such robustness allows Chern insulators to be fabricated into various device configurations without losing their remarkable dissipationless chiral edge states, making them promising platforms for low-power-consumption electronics [6,7]. Nonetheless, the comprehensive experimental verification of this robustness against geometric disruption is still lacking.

In recent years, intrinsic magnetic topological insulator $MnBi_2Te_4$ (MBT) has garnered significant interest in condensed matter physics [8,9]. As shown in Fig. 1(a), the crystal structure of MBT consists of stacked Te-Bi-Te-Mn-Te-Bi-Te septuple layers (SLs) [10]. Within each SL, the intralayer coupling of Mn ions establishes ferromagnetic order with an out-of-plane easy axis, while the antiferromagnetic exchange coupling between neighboring SLs results in an A-type antiferromagnetic ground state [11–17]. Crucially, in MBT, the Chern insulator state has been experimentally identified under both zero magnetic field and finite magnetic fields [18–24], characterized by a quantized Hall plateau and vanishing longitudinal resistance, which establishes MBT as an ideal platform for investigating emergent topological phenomena and engineering scalable quantum devices.

In this Letter, we systematically investigate the robustness of Chern insulator states in MBT thin flake devices against the geometric cuts introduced via atomic force microscopy (AFM). Initially, in pristine (uncut) devices, the $C = 1$ Chern insulator state under a moderate perpendicular magnetic field is observed, which is evidenced by a well-defined quantized Hall plateau and simultaneously vanishing longitudinal resistance. Subsequently, we introduce narrow structural cuts into the devices using an AFM tip and perform comprehensive transport characterizations using four-terminal, two-terminal, three-terminal, and nonlocal configurations. Crucially, the key topological transport properties, including quantized Hall resistance, vanishing longitudinal resistance and nearly dissipationless chiral transport, remain intact despite the severe geometric disruptions. Our Letter provides the first comprehensive experimental demonstration of the resilience of chiral edge states against structural defects, highlighting their topological protection and paving the way for the development of novel quantum topological devices based on robust chiral edge states.

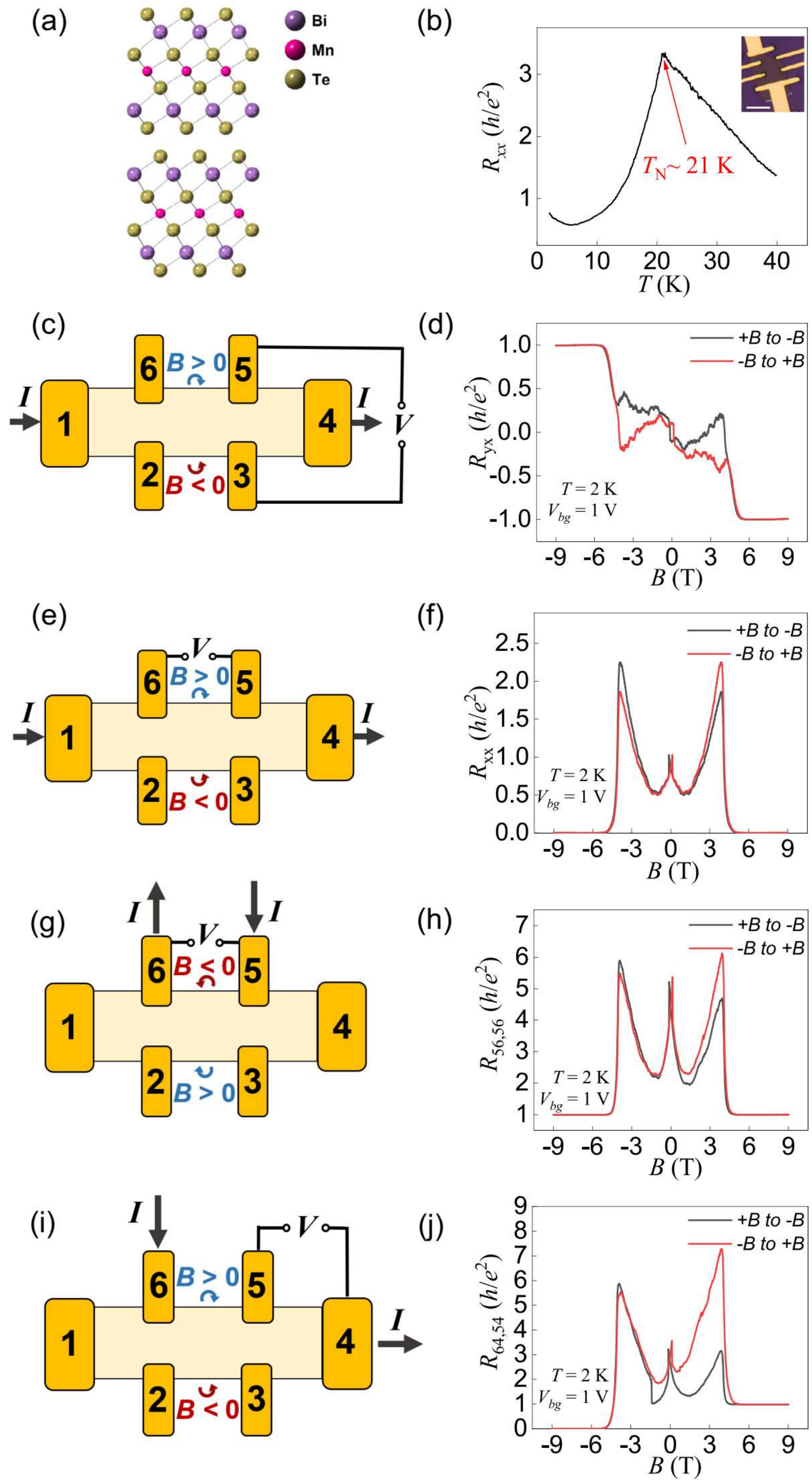

FIG. 1. Chern insulator states in $MnBi_2Te_4$ (MBT) device s1. (a) Schematic crystal structure of MBT. The pink, purple, and grayish-yellow spheres represent Mn, Bi, and Te atoms, respectively. (b) The temperature dependence of longitudinal resistance ($R_{xx}$)

for MBT device s1, revealing a Néel temperature of around 21 K. The inset shows the optical image of s1. The scale bar represents 10 μm. (c),(e) Schematics of the Hall-bar measurement configurations for (c) Hall (i.e., $R_{14,35}$) and (e) longitudinal resistance (i.e., $R_{14,65}$). The curved arrows indicate the chirality of the topological transport, which is clockwise for $B > 0$ and counterclockwise for $B < 0$. Here and hereafter, $R_{ij,kl}$ denotes a resistance measurement with the applied current flowing from lead $i$ to lead $j$, and voltage measured between leads $k$ ($V_+$) and $l$ ($V_-$). (d) Hall resistance ($R_{yx}$) and (f) $R_{xx}$ of device s1 as a function of magnetic field at $T = 2$ K with a backgate voltage of 1 V. The black and red traces represent magnetic field sweeping from positive to negative and from negative to positive directions, respectively. (g) Schematic of the two-terminal measurement configuration. The curved arrows indicate the chirality of the topological transport, which is clockwise for $B > 0$ and counterclockwise for $B < 0$. (h) Two-terminal resistance as a function of magnetic field at $T = 2$ K with a backgate voltage of 1 V. (i) Schematic of the three-terminal measurement configuration. The curved arrows indicate the chirality of the topological transport, which is clockwise for $B > 0$ and counterclockwise for $B < 0$. (j) Three-terminal resistance as a function of magnetic field at $T = 2$ K with a backgate voltage of 1 V.

The MBT thin flakes are mechanically exfoliated from high-quality MBT single crystals and transferred onto 285 nm-thick $SiO_2$/Si substrates. Subsequently, the thin flakes are fabricated into devices with standard Hall bar geometry. The inset of Fig. 1(b) displays the optical image of a 5-SL-thick MBT device s1 (see Supplemental Material for AFM results [25]). The temperature dependence of the longitudinal resistance ($R_{xx}$) shown in Fig. 1(b) exhibits an obvious resistance peak at around 21 K, indicative of the Néel temperature ($T_N$).

We systematically study the magnetotransport properties of the MBT devices. Figure 1 shows the results of the 5-SL MBT device s1. The quantized Hall resistance plateau of $0.996\,h/e^2$ is observed at $T = 2$ K and $|B| > 6$ T [Fig. 1(d)], accompanied by a nearly vanishing longitudinal resistance of $0.002\,h/e^2$ [Fig. 1(f)]. These features signify the realization of a Chern insulator state with Chern number $C = 1$. The magnetic phase transitions in MBT devices are shown in Appendix B. To further verify the robustness of the edge transport, we examine the two-terminal resistance of s1 [Fig. 1(g)]. As shown in Fig. 1(h), the measured two-terminal resistance $R_{56,56}$ exhibits a plateau of $0.996\,h/e^2$ in the quantized regime. Here and hereafter, $R_{ij,kl} = (V_k - V_l)/I_{ij}$, which denotes a resistance measurement with the applied current $I_{ij}$ flowing from lead $i$ to lead $j$, and the voltage measured between leads $k$ ($V_k$) and $l$ ($V_l$).

Theoretically, transport in the Chern insulator states is mediated by chiral edge channels that propagate unidirectionally without backscattering. Within the Landauer-Büttiker formalism [27,28], nearly dissipationless chiral transport implies near-unity transmission between adjacent contacts. Consequently, in a standard Hall bar geometry [Figs. 1(c) and 1(e)], the longitudinal resistance $R_{xx}$ is expected to vanish, while both the Hall resistance $R_{yx}$ and two-terminal resistance should be quantized to $h/e^2$ [1–3], fully consistent with our observations in Fig. 1. The directionality of the edge states can be further probed via three-terminal measurements, which are predicted to exhibit a chirality-dependent switching behavior. Consider a setup where current flows from a source lead to a drain lead, and the voltage is measured between an adjacent probe and the drain, as illustrated in Fig. 1(i). When the magnetic field directs the chiral edge states to propagate from the source directly toward the voltage probe, the probe equilibrates to the source potential, yielding a quantized resistance of $h/e^2$. Conversely, when the field is reversed, the edge states propagate in the opposite direction toward the drain. In this case, the probe equilibrates to the drain potential due to the absence of backscattering, resulting in a vanishing three-terminal resistance. Verifying these signatures is essential to confirm the establishment of robust chiral edge states.

This chirality-dependent switching behavior of the edge transport in our MBT device is explicitly captured by such three-terminal measurements. With current flowing from lead 6 to lead 4 [Fig. 1(i)], the three-terminal resistance $R_{64,54}$ displays a striking asymmetry: it is negligible ($< 1 \times 10^{-4}\, h/e^2$) for $B < -6$ T but robustly quantized ($0.985\, h/e^2$) for $B > 6$ T [Fig. 1(j)]. This behavior aligns perfectly with the theoretical picture described above: at $B < 0$, the counterclockwise edge mode renders the probe (lead 5) and the drain (lead 4) equipotential (vanishing $R_{64,54}$), whereas at $B > 0$, the reversed chirality forces the current to propagate to the probe initially, causing the probe to equilibrate with the source (lead 6) potential, resulting in a three-terminal resistance $R_{64,54} = h/e^2$. These results provide strong evidence for the chirality of the edge states in our MBT devices.

Having demonstrated the dissipationless chiral edge transport in the Chern insulator states of MBT, we now address a crucial question: can these quantum transport properties survive engineered geometric cuts? A positive confirmation would not only provide a direct demonstration of the topological protection inherent to Chern insulator states but also highlight their potential for constructing next-generation topological quantum devices [29,30].

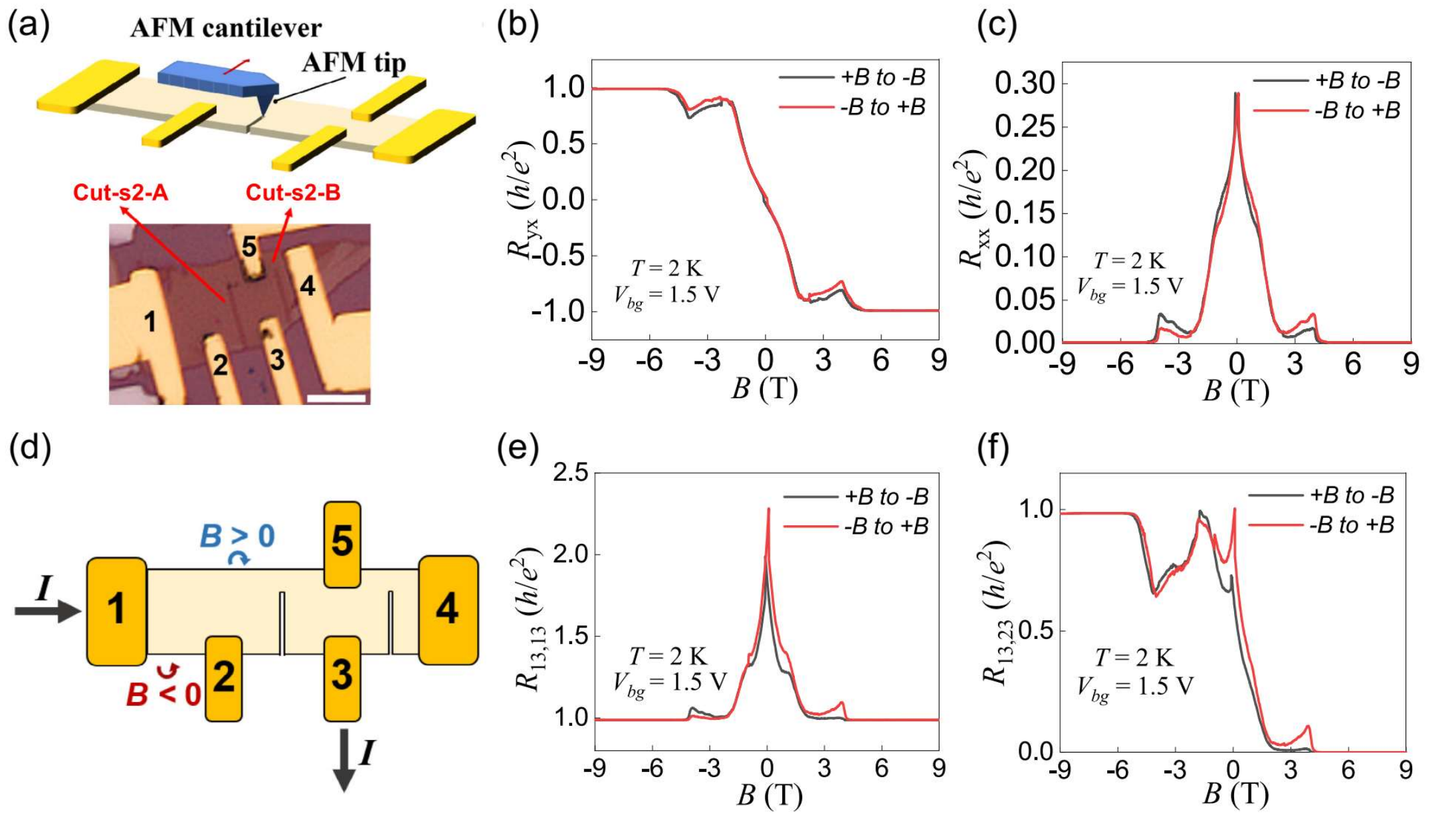

FIG. 2. Systematic magneto-measurements on MBT device s2 with engineered cuts. (a) Top: schematic of the AFM-based nanomachining setup used to create the cuts. Bottom: the optical image of device s2 with two long cuts. The positions of the two cuts are indicated by red arrows. The scale bar represents 10 μm. (b),(c) $R_{yx}$ (i.e., $R_{14,35}$) and $R_{xx}$ (i.e., $R_{14,23}$) of device s2 as a function of magnetic field at $T = 2$ K with a backgate voltage of 1.5 V. The black and red traces represent magnetic field sweeping from positive to negative and from negative to positive directions, respectively. (d) Schematic of the measurement configuration in MBT device s2 in the presence of two geometric cuts under $B > 0$ and $B < 0$, with the applied current flowing from lead 1 to lead 3. The curved arrows indicate the chirality of the topological transport, which is clockwise for $B > 0$ and counterclockwise for $B < 0$. (e),(f) Two-terminal and three-terminal resistance as a function of magnetic field at $T = 2$ K with a backgate voltage of 1.5 V. For both measurements, the applied current flows from lead 1 to lead 3.

To directly verify whether the topological transport can withstand geometric disruptions, we mechanically create cuts on the MBT devices using AFM nanomachining as a nanofabrication technique [31]. Specifically, the AFM tip was operated in contact mode with a significantly increased vertical loading force, allowing it to penetrate through the MBT flake. By precisely guiding the tip via piezoelectric actuators, the structural slits were created on the MBT flake (see Supplemental Material for more details [25]). This process, illustrated in the top panel of Fig. 2(a), enables the precisely controlled modification of the device geometry, providing a promising platform to examine the immunity of the chiral edge states to artificial cuts. To ensure the sample was completely cut through, we performed systematic characterizations on the cut. The AFM topography [Fig. S6(a) [25]] and high-resolution SEM images (Fig. S7 [25]) reveal a clean trench with no observable residue in the cut. More importantly, the AFM line profile shows no height step between the trench bottom and the bare substrate [Fig. S6(b) [25]], confirming that the cut reaches the substrate.

As shown in the bottom panel of Fig. 2(a), there are two linear cuts in device s2. Cut s2-A is located between leads 2 and 3, and cut s2-B is between leads 3 and 4. The two cuts are approximately 10 μm in length and hundreds of nanometers in width. Such a geometry is expected to severely disrupt diffusive bulk transport, which would be significantly scattered by such obstacles. To investigate the potential impact of the cuts on the topological transport, we conduct systematic measurements on the modified device s2. Remarkably, we observe a quantized Hall plateau with a Hall resistance $R_{yx}$ of around $0.99\,h/e^2$ at $T = 2$ K and $|B| > 6$ T, accompanied by a nearly vanishing longitudinal resistance $R_{xx}$ of around $0.001\,h/e^2$, as shown in Figs. 2(b) and 2(c). The observation of such nearly ideal quantization provides compelling evidence that the topological transport properties of chiral edge states are robust against geometric cuts.

The two-terminal measurements on s2, with the applied current flowing from lead 1 to lead 3 [Fig. 2(d)], exhibit a two-terminal resistance of $0.987\,h/e^2$ within the Chern insulator regime, serving as further verification of the nearly dissipationless transport [Fig. 2(e)]. To verify the persistence of chirality in the presence of the cuts, we further perform three-terminal measurements on s2 at $T = 2$ K. With the applied current flowing from lead 1 to lead 3 [Fig. 2(d)], vanishing three-terminal resistance ($R_{13,23} < 1 \times 10^{-4}\,h/e^2$) persists for $B > 6$ T, while a quantized plateau ($R_{13,23}$ of around $0.984\,h/e^2$) emerges for $B < -6$ T, which further confirms the chiral nature of the topological transport [Fig. 2(f)]. Taken together, the standard four-terminal, two-terminal, and three-terminal measurements provide strong evidence that the cuts neither introduce significant dissipation pathways nor disrupt the topological chiral nature.

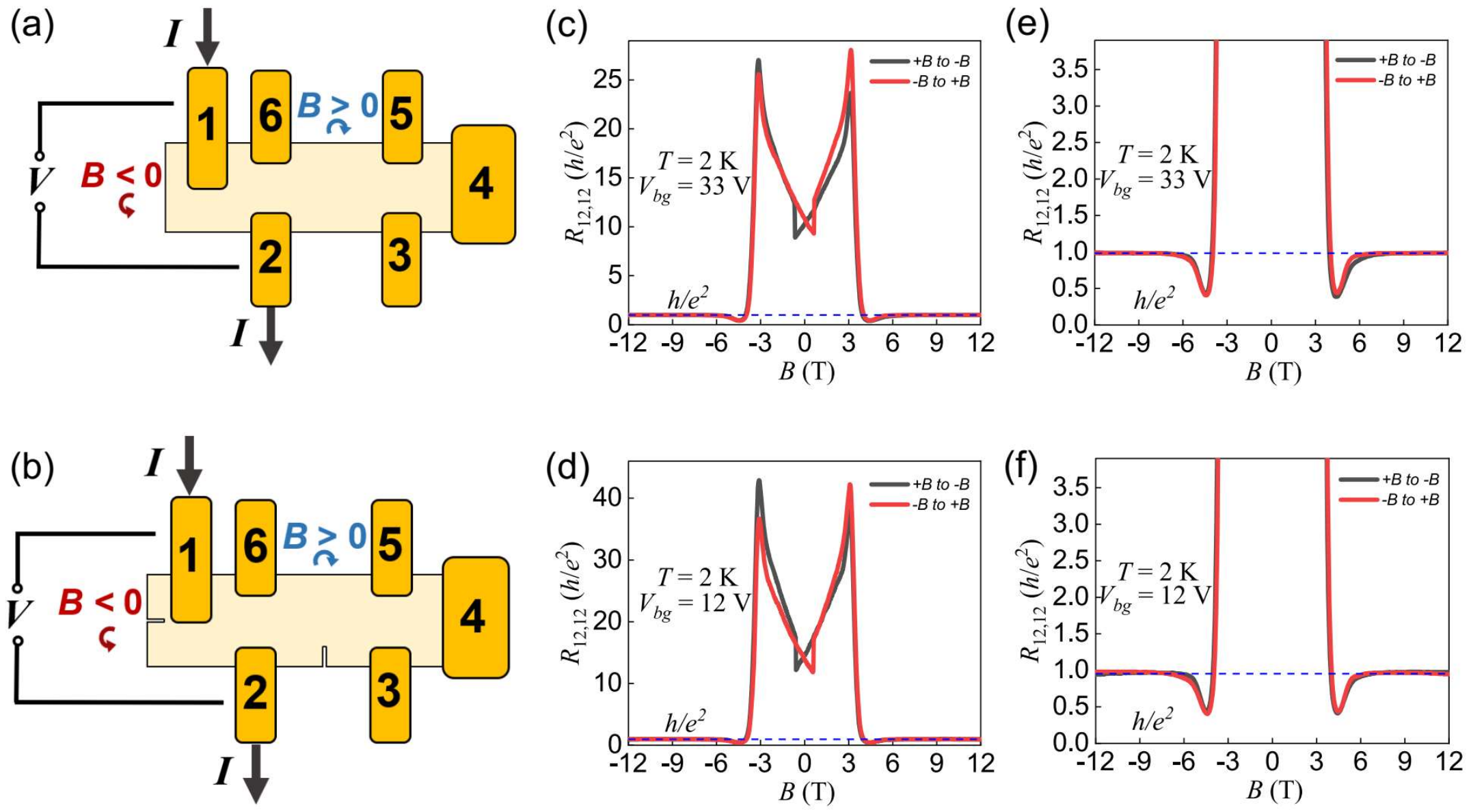

FIG. 3. Robustness of the quantized two-terminal resistance in device s3 before and after cutting. (a),(b) Schematics of the device configurations before (a) and after (b) introducing the two cuts. The curved arrows indicate the chirality of the topological transport, which is clockwise for $B > 0$ and counterclockwise for $B < 0$. (c) Two-

terminal resistance as a function of magnetic field on device s3 before cutting, measured at $T = 2$ K with a backgate voltage of 33 V. (d) Two-terminal resistance as a function of magnetic field on device s3 after cutting, measured at $T = 2$ K with a backgate voltage of 12 V. (e),(f) Enlargements of the plateaus in panels (c) and (d), respectively. The dashed lines denote the quantized value $h/e^2$.

As shown in Figs. 3(a) and 3(b), to directly investigate the impact of introduced structural cuts on the MBT device, we perform comparative two-terminal measurements on a 6-SL device s3 before and after the AFM cutting process (see Supplemental Material for optical images [25]). Two cuts (s3-A and s3-B) are introduced using the same AFM nanomachining technique employed for device s2. Cut s3-A is between lead 1 and 2 and cut s3-B is between lead 2 and 3. As shown in Figs. 3(c) and 3(e), before cutting, device s3 exhibits a two-terminal resistance of around $0.986\, h/e^2$ at 2 K under an external magnetic field $|B| > 6$ T and a backgate voltage of 33 V. After cutting, the quantized transport persists, with the two-terminal resistance reaching $\sim 0.980\, h/e^2$ for $|B| > 6$ T and a gate voltage of $V_g = 12$ V [Figs. 3(d) and 3(f)]. The difference in the operating backgate voltage may result from the Fermi level shift induced by environmental exposure during the *ex situ* AFM cutting process. The survival of the well-quantized resistance plateau, despite this invasive physical modification, provides strong evidence that the topological chiral edge states are immune to such engineered geometric deformations.

The chiral transport behavior can be described by the Landauer-Büttiker formalism, as discussed previously. The Landauer-Büttiker formalism connects the macroscopic electrical transport properties of a device to the microscopic quantum transmission probabilities ($T_{ij}$), serving as a powerful framework for analyzing the chiral transport characteristic in Chern insulators. In a multiterminal geometry, the net current $I_i$ flowing into the device through electrode $i$ is described by the relation $I_i = (\frac{e^2}{h})\sum_{i\neq j} T_{ij}(V_i - V_j)$, where $V_j$ is the voltage at electrode $j$, $T_{ij}$ is the transmission probability from electrode $j$ to $i$, and $I_i < 0$ denotes the net current flowing out of the device through electrode $i$. During the transport process dominated by chiral edge states, for $B < 0$, transmission is only allowed from an electrode $j$ to its immediate neighbor $i = j + 1$ (with indices taken cyclically), hence the transmission probability is $T_{ij} = \delta_{i,j+1}$ ($\delta_{i,j+1}$ is the Kronecker delta, defined as 1 if $i = j + 1$ and 0 otherwise). Conversely, for $B > 0$, the chirality is reversed, and the matrix becomes $T_{ij} = \delta_{i,j-1}$ [27,28].

Based on the Landauer-Büttiker formalism, Table I provides a comprehensive summary of the ideal resistance values for various measurement configurations in a standard six-terminal Hall bar geometry, using the lead notation defined in Fig. 1. These values, calculated by assuming ideal chiral edge transport, serve as a theoretical benchmark. The detailed derivations for the ideal values of such experimental setup are provided in the Appendix A.

The experimental results obtained from various measurement configurations on MBT devices with geometric cuts show excellent agreement with these theoretical predictions. As detailed in Table S1 [25], where comprehensive experimental data and quantitative consistency analyses are presented, the measured resistances align closely with the ideal values across all tested configurations. This excellent quantitative agreement confirms the robustness of the chiral edge states in MBT against geometric cuts, serving as a direct manifestation of their topological protection. Such immunity to structural disruptions is critical for practical applications, as defects are inevitable during the fabrication of scalable devices. Consequently, our findings highlight the Chern insulator as a promising platform for low-dissipation electronics, spintronics, and topological quantum information processing. Furthermore, these engineered cuts may serve as functional elements in their own right. For instance, it has been proposed that the proximity-effect-induced superconductivity in a Chern insulator with a cut could facilitate the emergence of Majorana zero modes at the cut terminals, paving the way for fault-tolerant topological quantum computing [29,30].

TABLE I. Theoretical resistance values for various measurement configurations based on ideal chiral edge transport. The electrode indices correspond to the Hall-bar geometry shown in Fig. 1.

| | Resistance ($h/e^2$) | | | Resistance ($h/e^2$) | | | Resistance ($h/e^2$) | |
|---|---|---|---|---|---|---|---|---|
| Configuration | $B > 0$ | $B < 0$ | Configuration | $B > 0$ | $B < 0$ | Configuration | $B > 0$ | $B < 0$ |
| $R_{14,14}$ | 1 | 1 | $R_{54,14}$ | 0 | 1 | $R_{64,14}$ | 0 | 1 |
| $R_{14,24}$ | 0 | 1 | $R_{54,24}$ | 0 | 1 | $R_{64,24}$ | 0 | 1 |
| $R_{14,34}$ | 0 | 1 | $R_{54,34}$ | 0 | 1 | $R_{64,34}$ | 0 | 1 |
| $R_{14,54}$ | 1 | 0 | $R_{54,54}$ | 1 | 1 | $R_{64,54}$ | 1 | 0 |
| $R_{14,64}$ | 1 | 0 | $R_{54,64}$ | 0 | 1 | $R_{64,64}$ | 1 | 1 |
| $R_{14,15}$ | 0 | 1 | $R_{54,15}$ | −1 | 0 | $R_{64,15}$ | −1 | 1 |
| $R_{14,25}$ | −1 | 1 | $R_{54,25}$ | −1 | 0 | $R_{64,25}$ | −1 | 1 |
| $R_{14,35}$ | −1 | 1 | $R_{54,35}$ | −1 | 0 | $R_{64,35}$ | −1 | 1 |
| $R_{14,65}$ | 0 | 0 | $R_{54,65}$ | −1 | 0 | $R_{64,65}$ | 0 | 1 |
| $R_{14,16}$ | 0 | 1 | $R_{54,16}$ | 0 | 0 | $R_{64,16}$ | −1 | 0 |
| $R_{14,26}$ | −1 | 1 | $R_{54,26}$ | 0 | 0 | $R_{64,26}$ | −1 | 0 |
| $R_{14,36}$ | −1 | 1 | $R_{54,36}$ | 0 | 0 | $R_{64,36}$ | −1 | 0 |
| $R_{14,21}$ | −1 | 0 | $R_{54,21}$ | 0 | 0 | $R_{64,21}$ | 0 | 0 |
| $R_{14,31}$ | −1 | 0 | $R_{54,31}$ | 0 | 0 | $R_{64,31}$ | 0 | 0 |
| $R_{14,32}$ | 0 | 0 | $R_{54,32}$ | 0 | 0 | $R_{64,32}$ | 0 | 0 |

We next discuss the possible microscopic transport pathways in these devices with cuts. Potential microscopic scenarios include conduction through possible residual MBT on the bottom of the trench, direct tunneling across the cut, rerouting of the chiral edge current along the newly created boundaries, and the coexistence of edge states with finite bulk conduction [32,33]. As established by our AFM and SEM characterizations discussed previously (Figs. S6 and S7 [25]), the structural cuts reach the substrate with no observable residual material.

To tunnel across the cuts, the chiral edge states must penetrate the vacuum barrier filling the cut. The tunneling probability across the cuts is determined by the external penetration depth ($\xi_{\mathrm{out}}$) of the edge states into the vacuum barrier. Specifically, the edge state wave functions are theoretically expected to decay exponentially into the vacuum with an amplitude proportional to $e^{-x/\xi_{\mathrm{out}}}$. Here, the external penetration depth is $\xi_{\mathrm{out}} = \hbar/\sqrt{2m\Phi}$, where $\Phi$ is the work function and $m$ is the free electron mass [34]. Considering that the work function $\Phi$ of MBT is about 4 eV [35,36], the external penetration depth is estimated to be about 0.1 nm, which is much smaller than the cut width in our MBT devices ranging from tens to hundreds of nanometers (Table S2 [25]). Therefore, direct vacuum tunneling across a clean trench is expected to be negligible.

As for the pure chiral transport along the reconstructed boundaries or the coexistence of edge transport with bulk conduction [32,33], our present transport results cannot make a clear distinction between these two scenarios. Nevertheless, such microscopic details do not affect our central conclusion that the topological transport response remains remarkably robust against severe engineered geometric disruptions.

In summary, we have demonstrated robust topological edge transport against geometric disruptions in the Chern insulator states of MBT devices. The $C = 1$ Chern insulator state, realized in MBT thin flake devices under a perpendicular magnetic field, remains robust after the introduction of mechanical cuts by AFM tips. Through various transport measurements including four-terminal, two-terminal, and three-terminal measurements, together with the direct comparison between the quantization results before and after the AFM cutting process in one single device, our Letter provides the first experimental evidence that chiral edge states in Chern insulators maintain topological protection even under extreme device structure modifications or damages, directly confirming their topological nature. This remarkable defect tolerance—demonstrated here in the MBT platform—establishes Chern insulators as a viable materials platform for implementing topologically robust quantum devices and scalable topological quantum circuits.

*Acknowledgments*—We thank K. T. Law for helpful discussions. This work was financially supported by the National Key Research and Development Program of China (Grants No. 2025YFA1411300 and No. 2022YFA1402404), the National Natural Science Foundation of China (Grants No. 12488201, No. 92565302, No. 12504158, No. 92565201, No. 124B2067, No. 12025404, No. 92161201, No. T2221003, and No. 12274208), Quantum Science and Technology-National Science and Technology Major Project (2021ZD0302403), the Natural Science Foundation of Jiangsu Province (Grants No. BK20230079, No. BK20243013, No. BK20233001), and the Fundamental and Interdisciplinary Disciplines Breakthrough Plan of the Ministry of Education of China (No. JYB2025XDXM411).

P. W., J. G., and J. L. contributed equally to this work.
*Contact author: junge@pku.edu.cn
†Contact author: jianwangphysics@pku.edu.cn

*Data availability*—The data that support the findings of this article are not publicly available. The data are available from the authors upon reasonable request.

## End Matter

*Appendix A: Detailed derivations of transporting properties by Landauer-Büttiker formalism*—Here, we present the step-by-step derivations for the ideal resistance values in different measurement configurations for s2 as an example, based on the assumption of a perfect chiral edge channel. The calculation begins with the Landauer-Büttiker formula, $I_i = (\frac{e^2}{h})\sum_{i\neq j}T_{ij}(V_i - V_j)$. We consider two cases for the transmission matrix $T_{ij}$: for a magnetic field $B < 0$, the edge channel propagates counterclockwise, giving $T_{ij} = \delta_{i,j+1}$. (The indices are cyclic, i.e., for a standard Hall bar with $N$ terminals, we identify the index $i = N + 1$ as $i = 1$. $\delta_{i,j+1}$ is the Kronecker delta, defined as 1 if $i = j + 1$ and 0 otherwise.) For $B > 0$, the propagation is clockwise, giving $T_{ij} = \delta_{i,j-1}$.

First, we derive the three-terminal resistance $R_{13,23}$ for the configuration where the applied current flows from lead 1 to lead 3 ($I_1 = I_0, I_3 = -I_0$). For the counterclockwise case ($B < 0$), the current at the drain, lead 3, is determined by $I_3 = (\frac{e^2}{h})T_{31}(V_3 - V_1) + (\frac{e^2}{h})T_{32}(V_3 - V_2) + (\frac{e^2}{h})T_{34}(V_3 - V_4) + (\frac{e^2}{h})T_{35}(V_3 - V_5)$. Considering $T_{32} = 1$ and $T_{31} = T_{34} = T_{35} = 0$, this expression reduces to $I_3 = [\frac{e^2}{h}](V_3 - V_2)$. Given that $I_3 = -I_0$, it follows that $-I_0 = [\frac{e^2}{h}](V_3 - V_2)$. The resistance $R_{13,23} = (V_2 - V_3)/I_0$ therefore simplifies to $h/e^2$.

Conversely, for the clockwise case ($B > 0$), we have $I_2 = (\frac{e^2}{h})T_{21}(V_2 - V_1) + (\frac{e^2}{h})T_{23}(V_2 - V_3) + (\frac{e^2}{h})T_{24}(V_2 - V_4) + (\frac{e^2}{h})T_{25}(V_2 - V_5) = 0$. Substituting $T_{23} = 1$ and $T_{21} = T_{24} = T_{25} = 0$ leads to $I_2 = (\frac{e^2}{h})T_{23}(V_2 - V_3) = 0$, which yields $V_2 = V_3$. The resistance $R_{13,23} = (V_2 - V_3)/I_0$ is thus exactly zero.

Next, we consider the two-terminal resistance $R_{13,13}$, where $I_1 = I_0$ and $I_3 = -I_0$. For the counterclockwise case ($B < 0$), the current at lead 3 is given by $I_3 = (\frac{e^2}{h})T_{31}(V_3 - V_1) + (\frac{e^2}{h})T_{32}(V_3 - V_2) + (\frac{e^2}{h})T_{34}(V_3 - V_4) + (\frac{e^2}{h})T_{35}(V_3 - V_5)$.

Applying $T_{32} = 1$ and $T_{31} = T_{34} = T_{35} = 0$, we have $I_3 = [\frac{e^2}{h}](V_3 - V_2) = -I_0$. Similarly, the current at lead 2 is $I_2 = (\frac{e^2}{h})T_{21}(V_2 - V_1) + (\frac{e^2}{h})T_{23}(V_2 - V_3) + (\frac{e^2}{h})T_{24}(V_2 - V_4) + (\frac{e^2}{h})T_{25}(V_2 - V_5) = 0$. Substituting $T_{21} = 1$ and $T_{23} = T_{24} = T_{25} = 0$, we obtain $I_2 = [\frac{e^2}{h}](V_2 - V_1) = 0$, which yields $V_2 = V_1$. Combining these results, we find $-I_0 = [\frac{e^2}{h}](V_3 - V_1)$. This yields $R_{13,13} = (V_1 - V_3)/I_0 = h/e^2$.

For the clockwise case ($B > 0$), the transmission coefficients are given by $T_{23} = 1$ and $T_{21} = T_{24} = T_{25} = 0$. Accordingly, $I_2 = (\frac{e^2}{h})T_{21}(V_2 - V_1) + (\frac{e^2}{h})T_{23}(V_2 - V_3) + (\frac{e^2}{h})T_{24}(V_2 - V_4) + (\frac{e^2}{h})T_{25}(V_2 - V_5) = [\frac{e^2}{h}](V_2 - V_3) = 0$, which yields $V_2 = V_3$. Since $T_{12} = 1$ and $T_{13} = T_{14} = T_{15} = 0$, the current at lead 1 becomes $I_1 = (\frac{e^2}{h})T_{12}(V_1 - V_2) + (\frac{e^2}{h})T_{13}(V_1 - V_3) + (\frac{e^2}{h})T_{14}(V_1 - V_4) + (\frac{e^2}{h})T_{15}(V_1 - V_5) = [\frac{e^2}{h}](V_1 - V_2)$. Substituting $I_1 = I_0$ and $V_2 = V_3$, we again find $I_0 = [\frac{e^2}{h}](V_1 - V_3)$, yielding $R_{13,13} = h/e^2$. Thus, the two-terminal resistance is quantized regardless of chirality.

Finally, we derive the four-terminal Hall resistance $R_{14,35}$ for the configuration where current $I_0$ is injected at lead 1 and drained at lead 4 ($I_1 = I_0, I_4 = -I_0$). We first consider the counterclockwise case ($B < 0$), where the transmission coefficients are given by $T_{21} = T_{32} = 1$ and $T_{23} = T_{24} = T_{25} = T_{31} = T_{34} = T_{35} = 0$. The zero-current conditions for the leads 2 and 3 thus yield $I_2 = (\frac{e^2}{h})T_{21}(V_2 - V_1) + (\frac{e^2}{h})T_{23}(V_2 - V_3) + (\frac{e^2}{h})T_{24}(V_2 - V_4) + (\frac{e^2}{h})T_{25}(V_2 - V_5) = [\frac{e^2}{h}](V_2 - V_1) = 0$ and $I_3 = (\frac{e^2}{h})T_{31}(V_3 - V_1) + (\frac{e^2}{h})T_{32}(V_3 - V_2) + (\frac{e^2}{h})T_{34}(V_3 - V_4) + (\frac{e^2}{h})T_{35}(V_3 - V_5) = [\frac{e^2}{h}](V_3 - V_2) = 0$. These equations together yield $V_3 = V_2 = V_1$. For lead 5, transmission coefficient is $T_{54} = 1$ and $T_{51} = T_{52} = T_{53} = 0$ and the zero-current condition is thus $I_5 = (\frac{e^2}{h})T_{51}(V_5 - V_1) + (\frac{e^2}{h})T_{52}(V_5 - V_2) + (\frac{e^2}{h})T_{53}(V_5 - V_3) + (\frac{e^2}{h})T_{54}(V_5 - V_4) = [\frac{e^2}{h}](V_5 - V_4) = 0$, which gives $V_5 = V_4$. The total current $I_0$ injected at the source gives $I_1 = (\frac{e^2}{h})T_{12}(V_1 - V_2) + (\frac{e^2}{h})T_{13}(V_1 - V_3) + (\frac{e^2}{h})T_{14}(V_1 - V_4) + (\frac{e^2}{h})T_{15}(V_1 - V_5) = [\frac{e^2}{h}](V_1 - V_5) = I_0$. Note that following the cyclic indexing mentioned above, $V_{N+1}$ is identified as $V_1$ for this Hall bar structure with $N = 5$ terminals (Fig. 2). Since we have already established $V_3 = V_2 = V_1$ and $V_5 = V_4$, the Hall resistance is then calculated as $R_{14,35} = (V_3 - V_5)/I_0 = (V_1 - V_5)/\{[\frac{e^2}{h}](V_1 - V_5)\} = h/e^2$.

For the clockwise case ($B > 0$), since $T_{12} = T_{23} = T_{34} = T_{45} = T_{51} = 1$ and other transmission coefficients are zero, similarly we have $I_3 = (\frac{e^2}{h})T_{31}(V_3 - V_1) + (\frac{e^2}{h})T_{32}(V_3 - V_2) + (\frac{e^2}{h})T_{34}(V_3 - V_4) + (\frac{e^2}{h})T_{35}(V_3 - V_5) = (\frac{e^2}{h})(V_3 - V_4) = 0$; $I_2 = (\frac{e^2}{h})T_{21}(V_2 - V_1) + (\frac{e^2}{h})T_{23}(V_2 - V_3) + (\frac{e^2}{h})T_{24}(V_2 - V_4) + (\frac{e^2}{h})T_{25}(V_2 - V_5) = [\frac{e^2}{h}](V_2 - V_3) = 0$ and $I_5 = (\frac{e^2}{h})T_{51}(V_5 - V_1) + (\frac{e^2}{h})T_{52}(V_5 - V_2) + (\frac{e^2}{h})T_{53}(V_5 - V_3) + (\frac{e^2}{h})T_{54}(V_5 - V_4) = [\frac{e^2}{h}](V_5 - V_1) = 0$, so $V_3 = V_2 = V_4$ and $V_5 = V_1$. Note that following the cyclic indexing mentioned above, $V_{N+1}$ is identified as $V_1$ for this Hall bar structure with $N = 5$ terminals (Fig. 2). The current at lead 1 is $I_1 = (\frac{e^2}{h})T_{12}(V_1 - V_2) + (\frac{e^2}{h})T_{13}(V_1 - V_3) + (\frac{e^2}{h})T_{14}(V_1 - V_4) + (\frac{e^2}{h})T_{15}(V_1 - V_5) = [\frac{e^2}{h}](V_1 - V_2) = I_0$. The Hall resistance is then $R_{14,35} = (V_3 - V_5)/I_0 = (V_2 - V_1)/\{[\frac{e^2}{h}](V_1 - V_2)\} = -h/e^2$. Thus, the magnitude of the Hall resistance is robustly quantized at $h/e^2$, independent of the field direction.

*Appendix B: Magnetic phase transitions in MBT devices*—At zero magnetic field, the MBT devices are in the A-type antiferromagnetic state. As the magnetic field increases from zero, the devices transition from the antiferromagnetic state to a canted antiferromagnetic state at $B_{\mathrm{AFM}}$ [20,24,37], identified as the peak positions in the $R_{xx}$-$B$ curves [Figs. 4(b) and 4(d)]. Further increasing the magnetic field induces a second transition at $B_{\mathrm{FM}}$. At this point, the magnetic moments fully align, driving the system into a ferromagnetic state, in which the quantized Hall resistance plateau [Figs. 4(a) and 4(c)] and nearly vanishing longitudinal resistance [Figs. 4(b) and 4(d)] are observed.

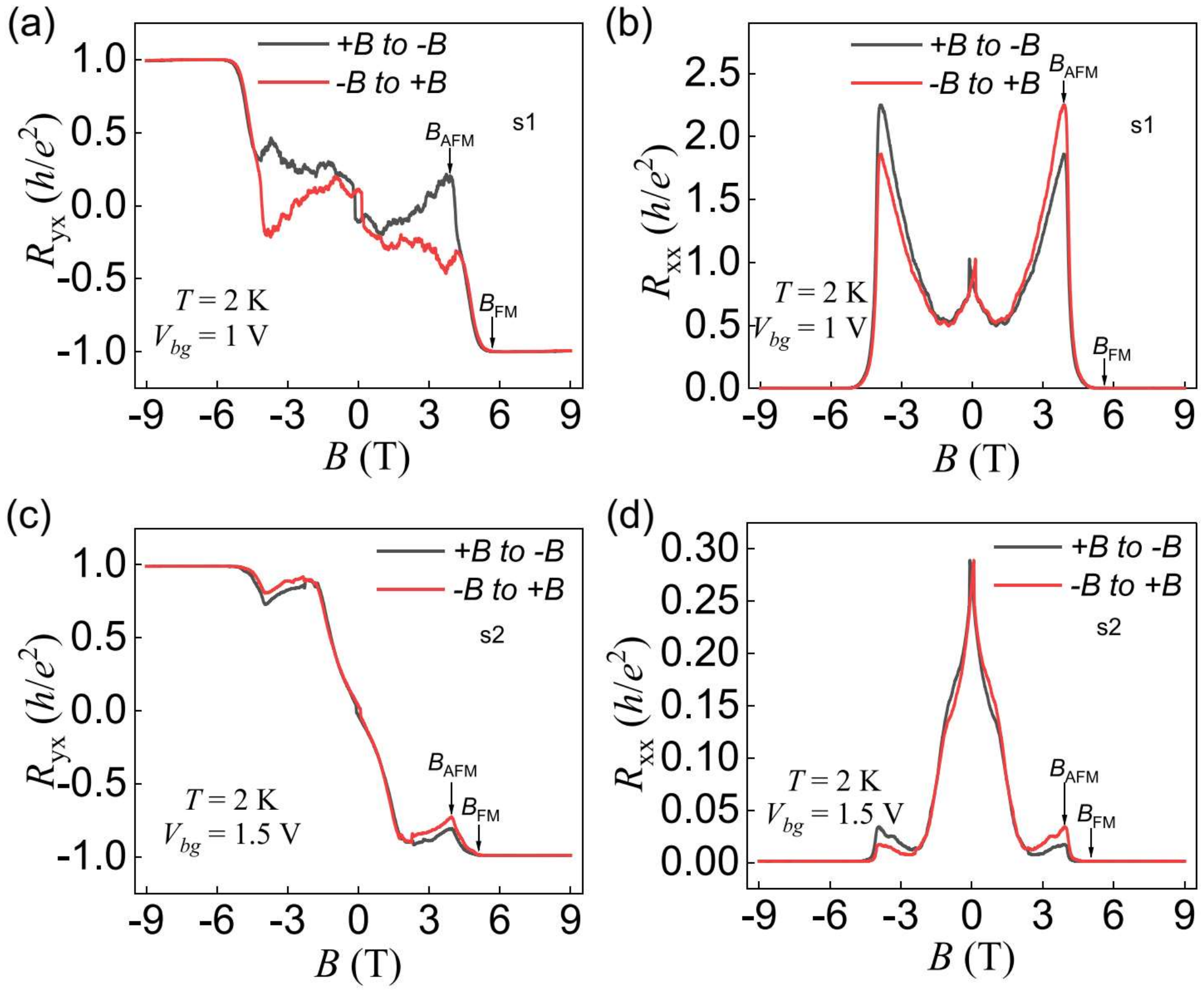

FIG. 4. Magnetic field dependence of the $R_{yx}$ and $R_{xx}$ in MBT devices s1 and s2. (a) $R_{yx}$ and (b) $R_{xx}$ as a function of magnetic field for s1. (c) $R_{yx}$ and (d) $R_{xx}$ as a function of magnetic field for s2. The critical magnetic fields for the spin-flop transition ($B_{\mathrm{AFM}}$) and the saturation field ($B_{\mathrm{FM}}$) are indicated by black arrows.

# Supplemental Material for "Demonstration of robust chiral edge transport in field-induced Chern insulator $MnBi_2Te_4$ devices with engineered geometric defects"

Pinyuan Wang[1,2], Jun Ge[1,2†], Jiawei Luo[3], Xiaoqi Liu[1,2], Jiayi Yang[1,2], Fucong Fei[4], Fengqi Song[5,6], & Jian Wang[1,2,7,8*]

[1]*International Center for Quantum Materials, School of Physics, Peking University, Beijing 100871, China*
[2]*Beijing Key Laboratory of Quantum Devices, Peking University, Beijing 100871, China*
[3]*State Key Laboratory of Quantum Functional Materials, and ShanghaiTech Laboratory for Topological Physics, School of Physical Science and Technology, ShanghaiTech University, Shanghai 201210, China*
[4]*National Laboratory of Solid State Microstructures, Collaborative Innovation Center of Advanced Microstructures, and School of Advanced Manufacturing Engineering, Nanjing University, Suzhou 215163, China*
[5]*National Laboratory of Solid State Microstructures, Collaborative Innovation Center of Advanced Microstructures, and School of Physics, Nanjing University, Nanjing 210093, China*
[6]*Atom Manufacturing Institute, Nanjing 211806, China*
[7]*Hefei National Laboratory, Hefei 230088, China*
[8]*Collaborative Innovation Center of Quantum Matter, Beijing 100871, China*

P.W., J.G. and J.L. contributed equally to this work.
*Corresponding author.
Jian Wang (jianwangphysics@pku.edu.cn)
†Corresponding author.
Jun Ge (junge@pku.edu.cn)

## Contents

## I. Materials and Methods

**Crystal Growth.** $MnBi_2Te_4$ (MBT) single crystals were grown using the flux method [1]. Precursors of MnTe and $Bi_2Te_3$ were mixed in a molar ratio of 1:5.85 and loaded into an alumina crucible sealed within a vacuum quartz ampule. Subsequently, the growth process was performed by heating the mixture to 950 °C over 24 h, maintaining it for 12 h, and cooling to 580 °C at a rate of 10 °C/h. Finally, the excess $Bi_2Te_3$ flux was removed by centrifugation to obtain the single crystals.

**Device Fabrication and Atomic Force Microscopy (AFM) Nanomachining.** Thin flakes of $MnBi_2Te_4$ were mechanically exfoliated from high-quality single crystals using scotch tape and transferred onto 285 nm-thick $SiO_2$/Si substrates. Prior to transfer, the substrates were pre-cleaned in oxygen plasma for five minutes at a pressure of approximately 80 mTorr. To obtain flakes with thicknesses down to several nanometers, the substrate was heated at 373 K (100 °C) for one minute after covering with the scotch tape. After spin-coating PMMA resist, standard electron beam lithography was performed using a FEI Helios NanoLab 600i Dual Beam system to define the electrode patterns. After *in-situ* Ar plasma cleaning to remove residual resist, metal contacts (Ti/Au, 6.5/150 nm) were deposited on the thin flake in a LJUHV E-400L E-Beam Evaporator, followed by a standard lift-off process.

To introduce geometric damages for testing edge state robustness, we performed AFM nanomachining on the MBT flakes using a Bruker DimensionICON system to create structural cuts. The AFM-tip was first engaged on the sample surface, and the vertical loading force was subsequently increased to a level sufficient to mechanically penetrate the MBT flake. Guided by the piezoelectric scanner, the tip was moved along pre-defined paths to physically remove material, creating structural cuts that reach the substrate with no observable residual material.

**Transport Measurements.** Electrical transport measurements were conducted in Physical Property Measurement Systems (re-liquefier based 16 T PPMS for device s3; cryogen-free PPMS Dynacool 9 T for devices s1, s2 and s4). The resistance was measured using Stanford Research Systems SR830 lock-in amplifiers. An AC excitation current of 100 nA with a frequency of 13.333 Hz was applied for both local and non-local measurement configurations.

**AFM and Scanning Electron Microscopy (SEM) characterizations.** The topography and the thickness of the samples were characterized using a Bruker DimensionICON AFM system. The geometric details of the samples were examined using a FEI Helios NanoLab 600i Dual Beam system.

## II. Supplemental Figures and Table

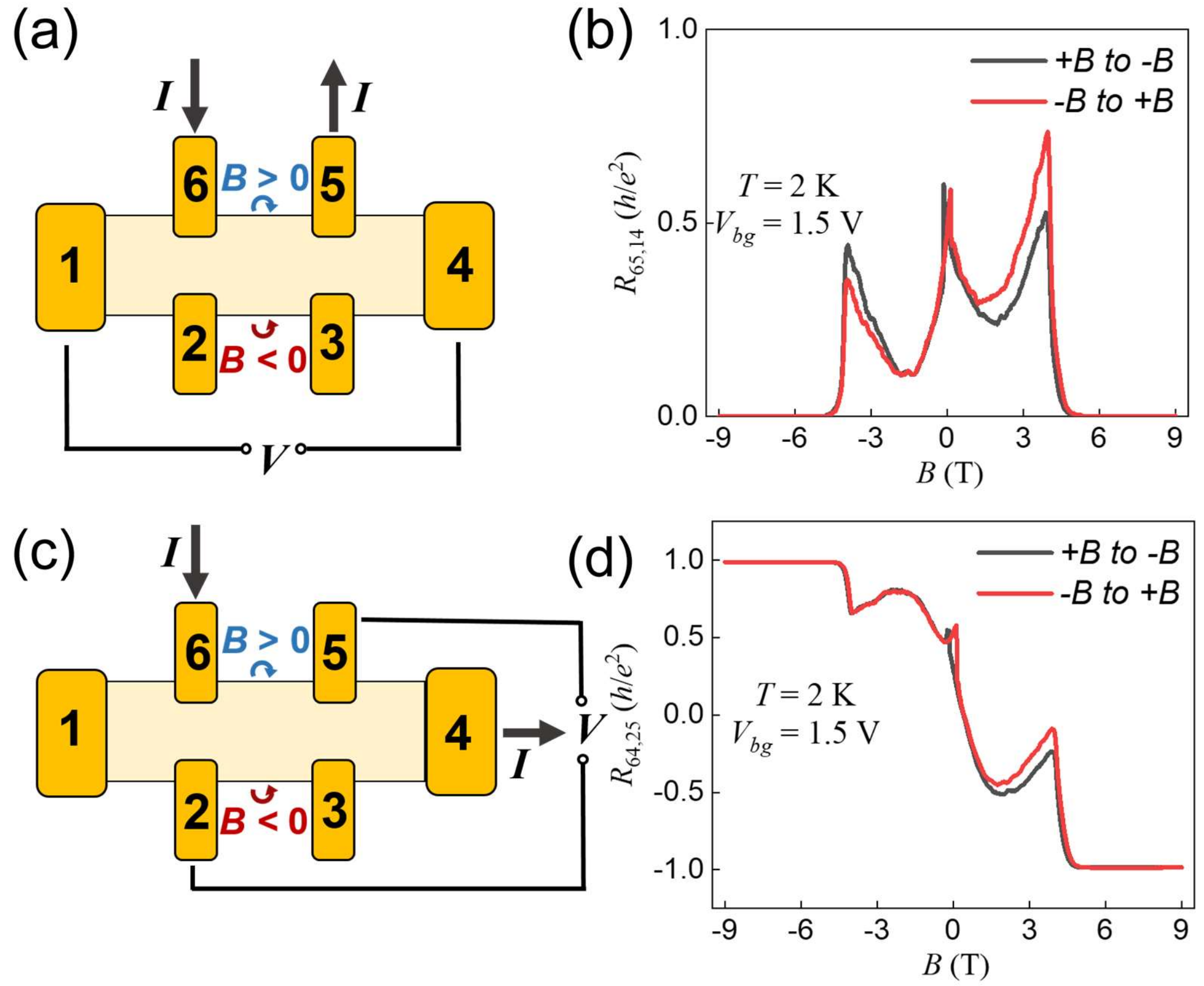

FIG. S1. Non-local measurements on MBT device s1 revealing the emergence of Chern insulator behavior under moderate perpendicular magnetic fields. (a) Schematic of the device configuration for the non-local measurement with the applied current flowing from lead 6 to 5. The curved arrows indicate the chirality of the transport, which is clockwise for $B > 0$ and counterclockwise for $B < 0$. (b) Magnetic field dependence of the nonlocal resistance $R_{65,14}$ of s1 taken at $T = 2$ K. The non-local resistance $R_{65,14}$ shows nearly vanishing values at $|B| > 6$ T, indicating dissipationless transport. (c) Schematic of the device configuration for the non-local measurements for the current flowing from lead 6 to lead 4. The curved arrows indicate the chirality of the topological transport, which is clockwise for $B > 0$ and counterclockwise for $B < 0$. (d) Magnetic field dependence of the nonlocal resistance $R_{64,25}$ of s1 taken at $T = 2$ K. The non-local resistance $R_{64,25}$ becomes quantized to 0.988 $h/e^2$ at $|B| > 6$ T. These results confirm the robustness of the chiral edge states.

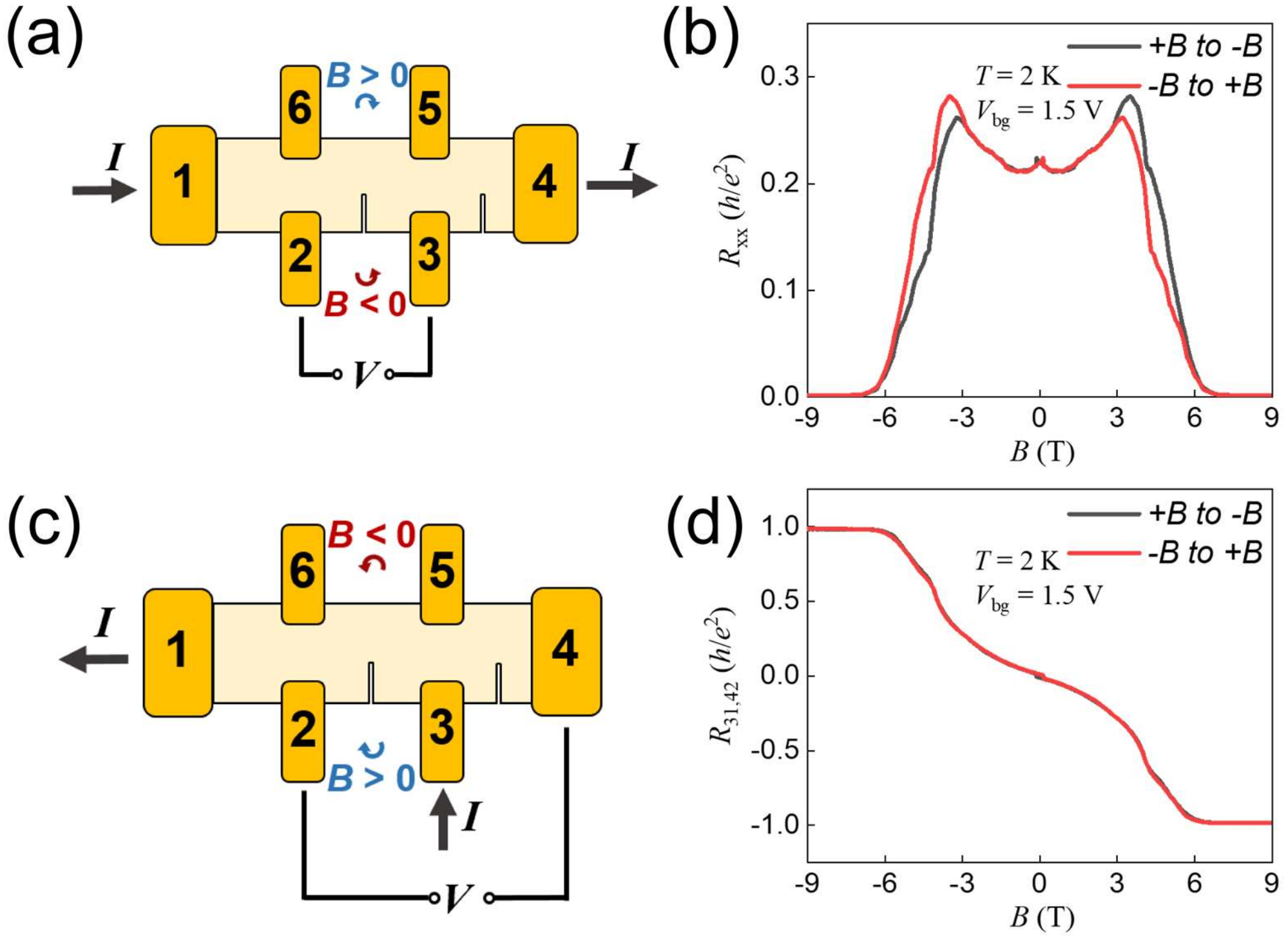

FIG. S2. Electrical transport measurements on a 5-SL MBT device s4 with cuts. There are two linear cuts in device s4. Cut s4-A is located between leads 2 and 3, and cut s4-B is between leads 3 and 4. (a) Schematic of the measurement configuration for longitudinal resistance (i.e. $R_{14,23}$), the curved arrows indicate the chirality of the topological transport, which is clockwise for $B > 0$ and counterclockwise for $B < 0$. The applied current flows from lead 1 to lead 4. (b) Magnetic field dependence of the longitudinal resistance of s4 taken at $T = 2$ K with a backgate voltage of 1.5 V. The longitudinal resistance exhibits nearly vanishing values at $|B| > 8$ T, indicating dissipationless transport, consistent with the behavior of a Chern insulator despite the presence of the cuts. (c) Schematic of the device configuration for non-local measurements. The applied current flows from lead 3 to lead 1. The curved arrows indicate the chirality of the topological transport, which is clockwise for $B > 0$ and counterclockwise for $B < 0$. (d) Magnetic field dependence of the nonlocal resistance $R_{31,42}$ of s4 taken at $T = 2$ K with a backgate voltage of 1.5 V. The non-local resistance $R_{31,42}$ reaches nearly quantized values of 0.985 $h/e^2$ at $|B| > 8$ T, confirming the stability of the chiral edge states. These results demonstrate that the Chern insulator state in device s4 remains robust after the AFM cuts were introduced.

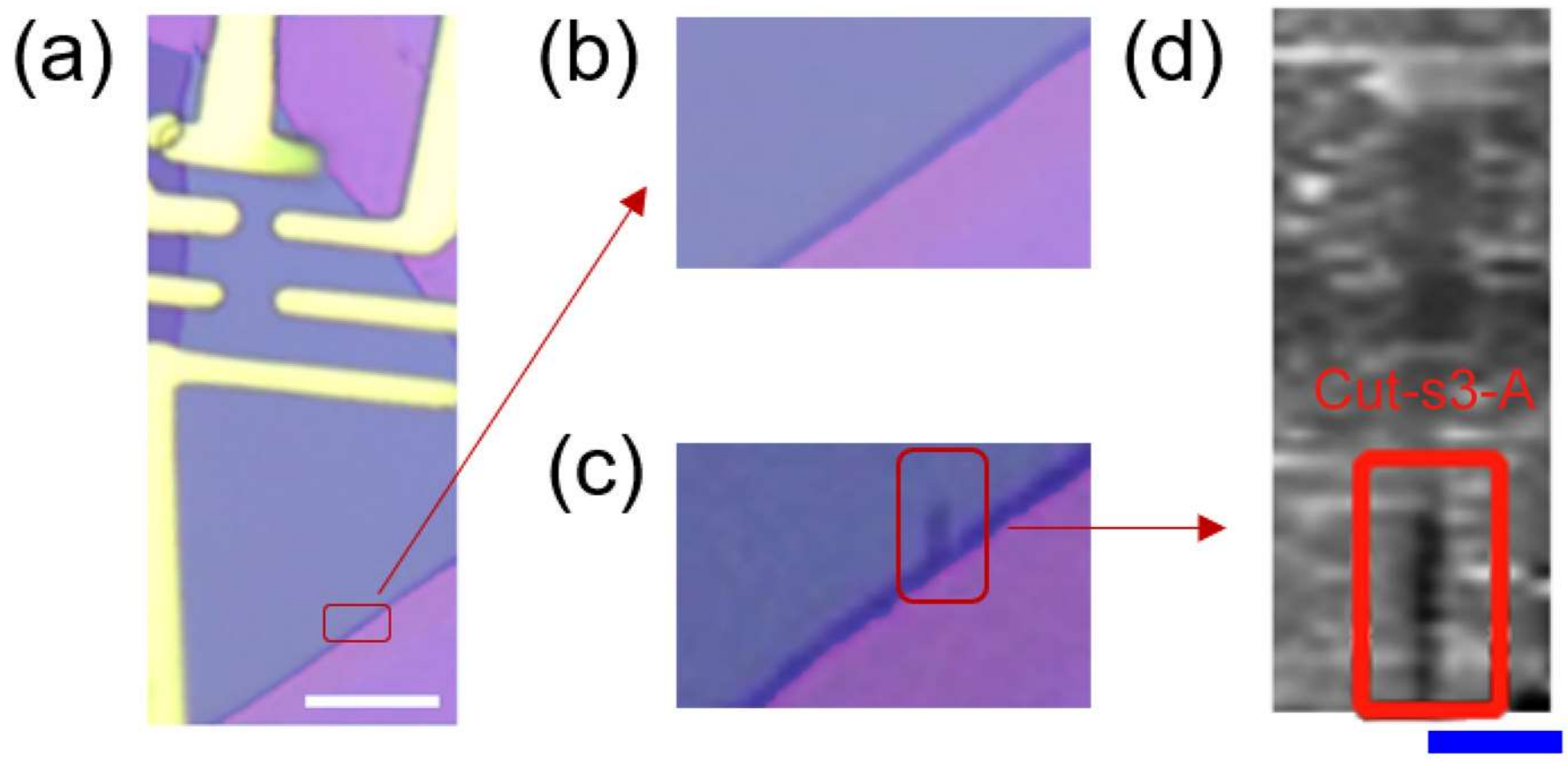

FIG. S3. False-colored optical image and AFM image of MBT device s3 before and after AFM nanomachining. (a) False-colored optical image of the pristine device s3. The scale bar (white) represents 10 μm. (b) Enlarged view of the selected area of the pristine device s3 in (a). (c) False-colored optical image of the same area of the device shown in (b) after introducing the geometric defect (Cut s3-A). The red arrow indicates the position of the cut. (d) AFM topography of the cut region, corresponding to the area marked by the red rectangle. The length of the cut is estimated to be about 700 nm. The scale bar (blue) represents 500 nm.

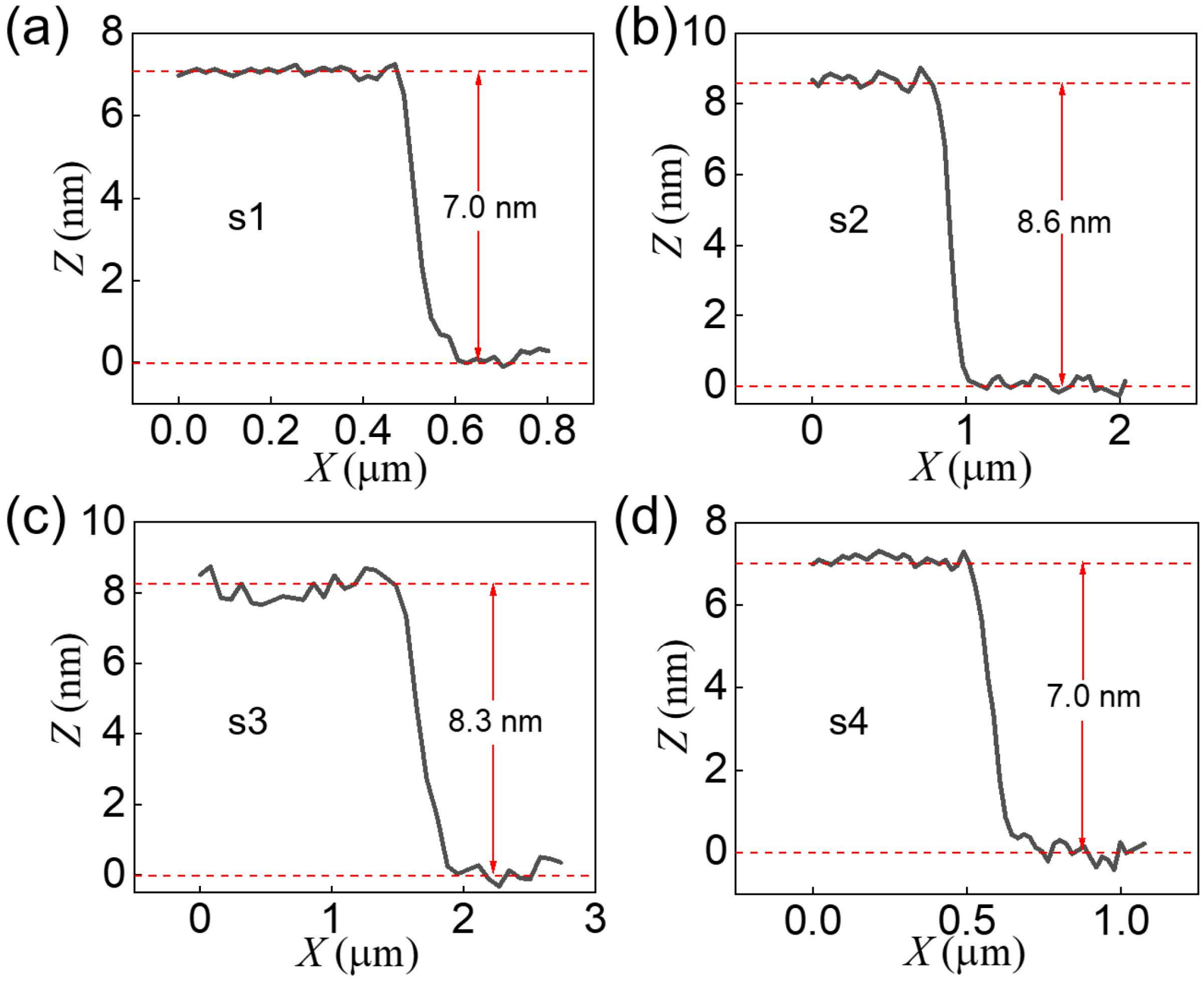

FIG. S4. Thickness of MBT device s1 (a), s2 (b), s3 (c) and s4 (d) measured by AFM.

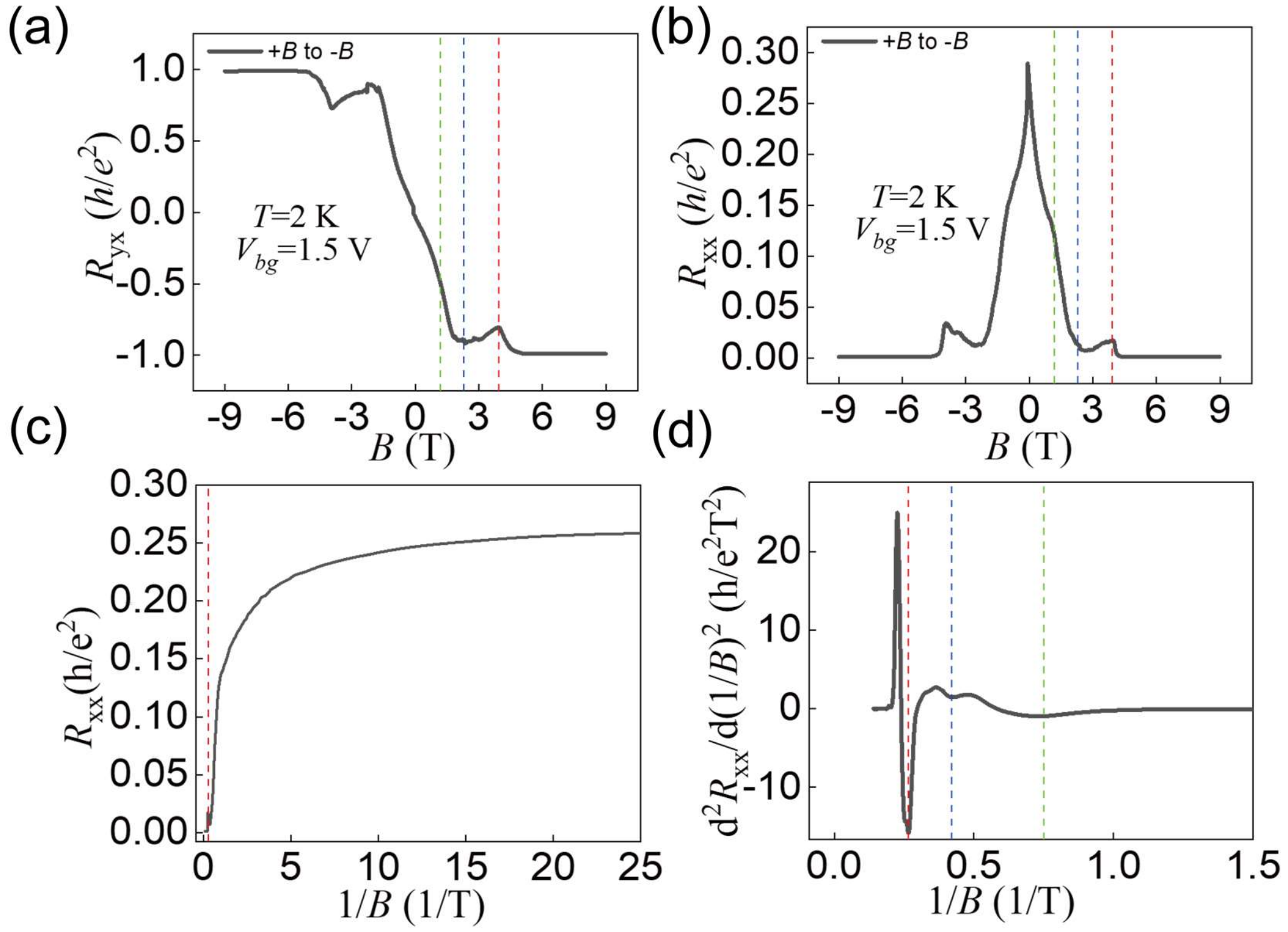

FIG. S5. Absence of Shubnikov-de Haas (SdH) oscillations in our MBT device. (a) $R_{yx}$ plotted as a function of the magnetic field in MBT device s2. (b) $R_{xx}$ plotted as a function of magnetic field in MBT device s2. (c) $R_{xx}$ plotted as a function of the inverse magnetic field (1/$B$). No oscillations are observed in the $R_{xx}$ curve in MBT device s2. The kink at around 3.9 T marked by the red dashed line corresponds to the transition into the canted antiferromagnetic state [2-5]. (d) The second derivative of the longitudinal resistance with respect to the inverse magnetic field, $d^2R_{xx}/d(1/B)^2$. No periodic oscillations are observed in the second-derivative curve. The three non-periodic structures at around 3.9 T, 2.3 T and 1.2 T (marked by the red, blue and green dashed lines, respectively) in the second-derivative curve correspond to the small kinks in $R_{xx}$ (also marked by the red, blue and green dashed lines in (b) and (a)). The structure at 3.9 T corresponds to the aforementioned magnetic transition into the canted antiferromagnetic state. The minor features at 2.3 T and 1.2 T may originate from spin flopping during the magnetization process [5].

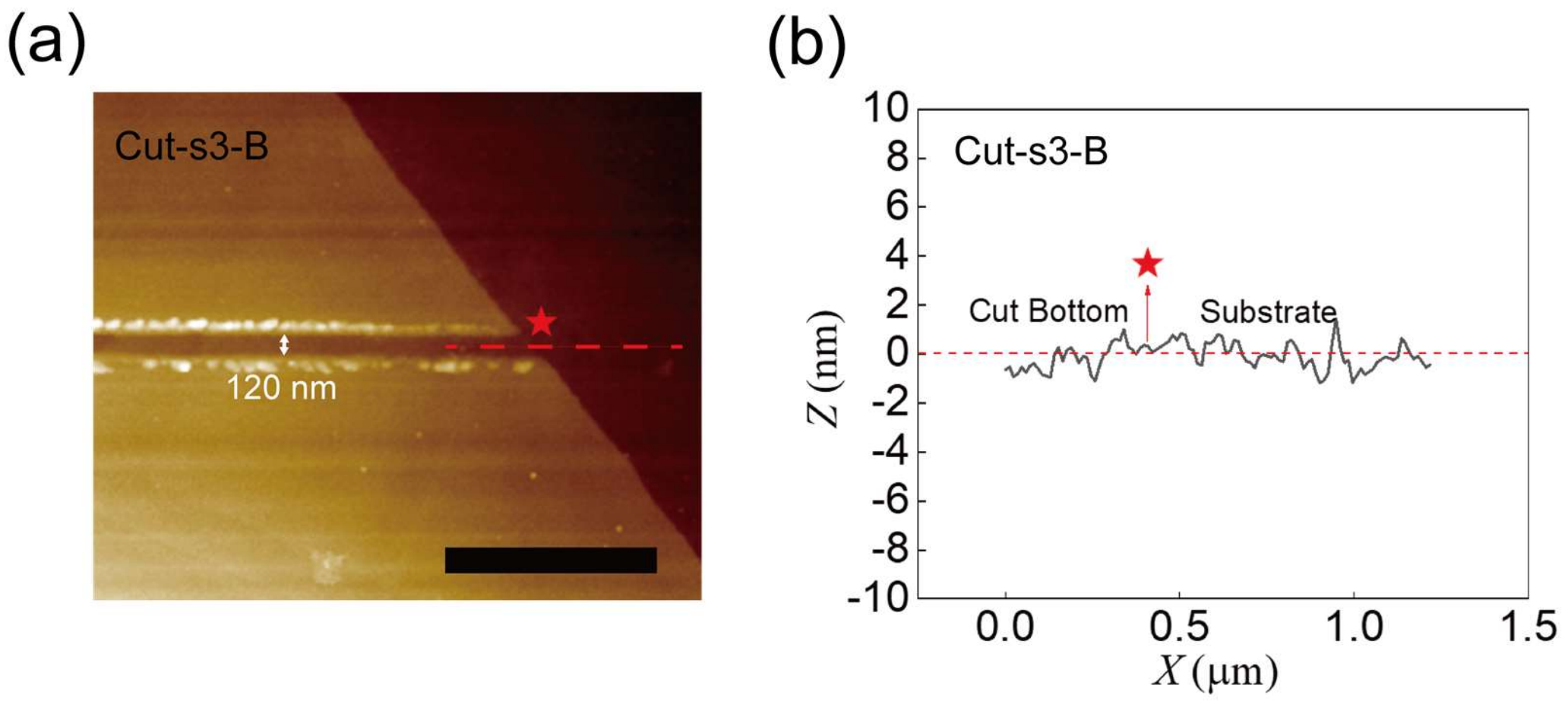

FIG. S6. AFM characterizations on cut s3-B in MBT device s3. (a) High-resolution AFM topography image of the cut region. The light-colored region on the left is the MBT thin flake, and the dark-colored region on the right is the $SiO_2$/Si substrate. The star symbol indicates the boundary between the bottom of the cut and the substrate. The scale bar represents 1 μm. (b) AFM line profile taken along the red dashed line in (a), from the bottom of the cut extending to the bare substrate. The star symbol marks the boundary position corresponding to that in (a). The absence of a height step between the trench bottom and the bare substrate confirms that the cut penetrates down to the substrate.

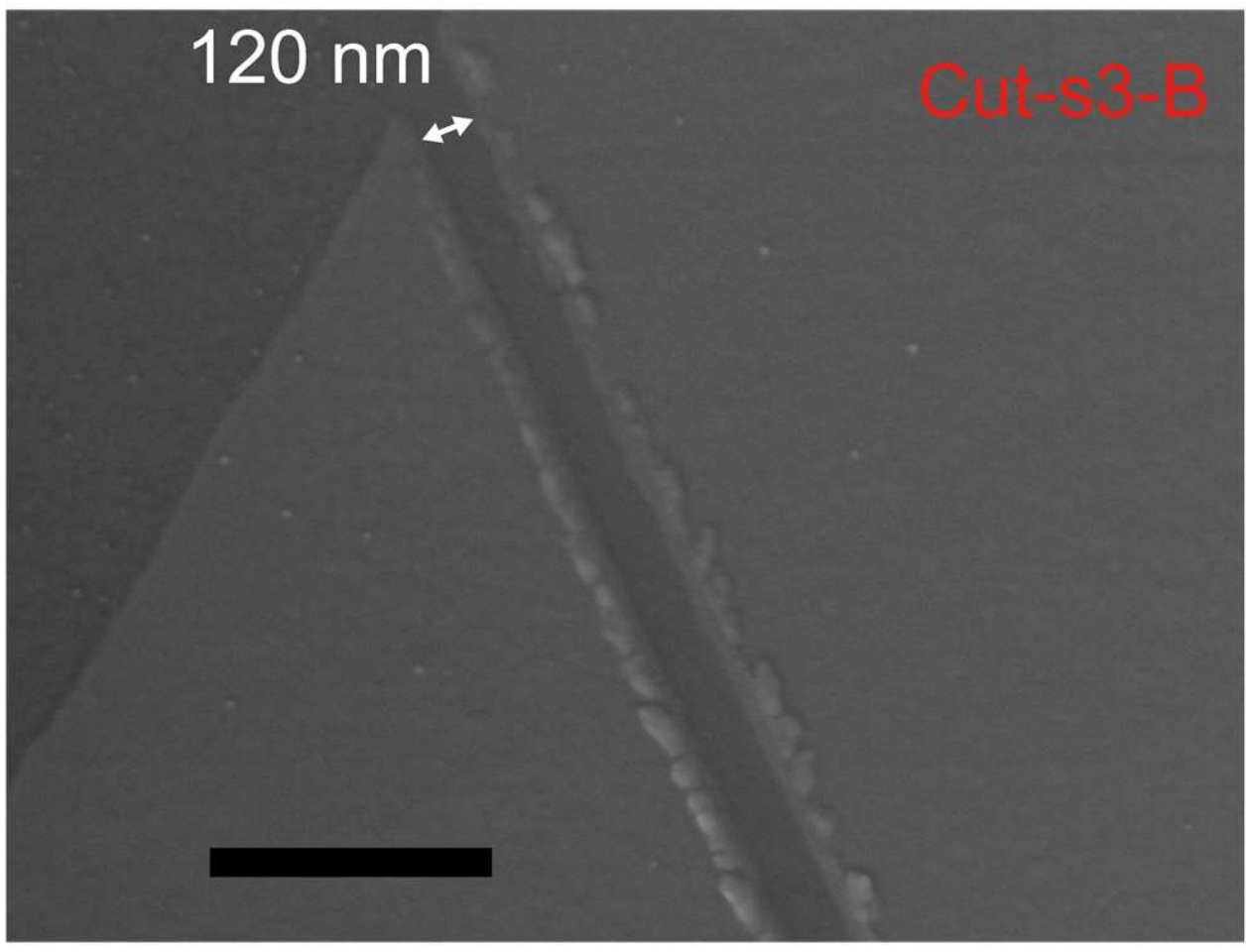

FIG. S7. SEM image of cut s3-B in device s3. The image reveals a cut with a width of approximately 120 nm, showing no observable sample residue in the trench. The scale bar represents 500 nm.

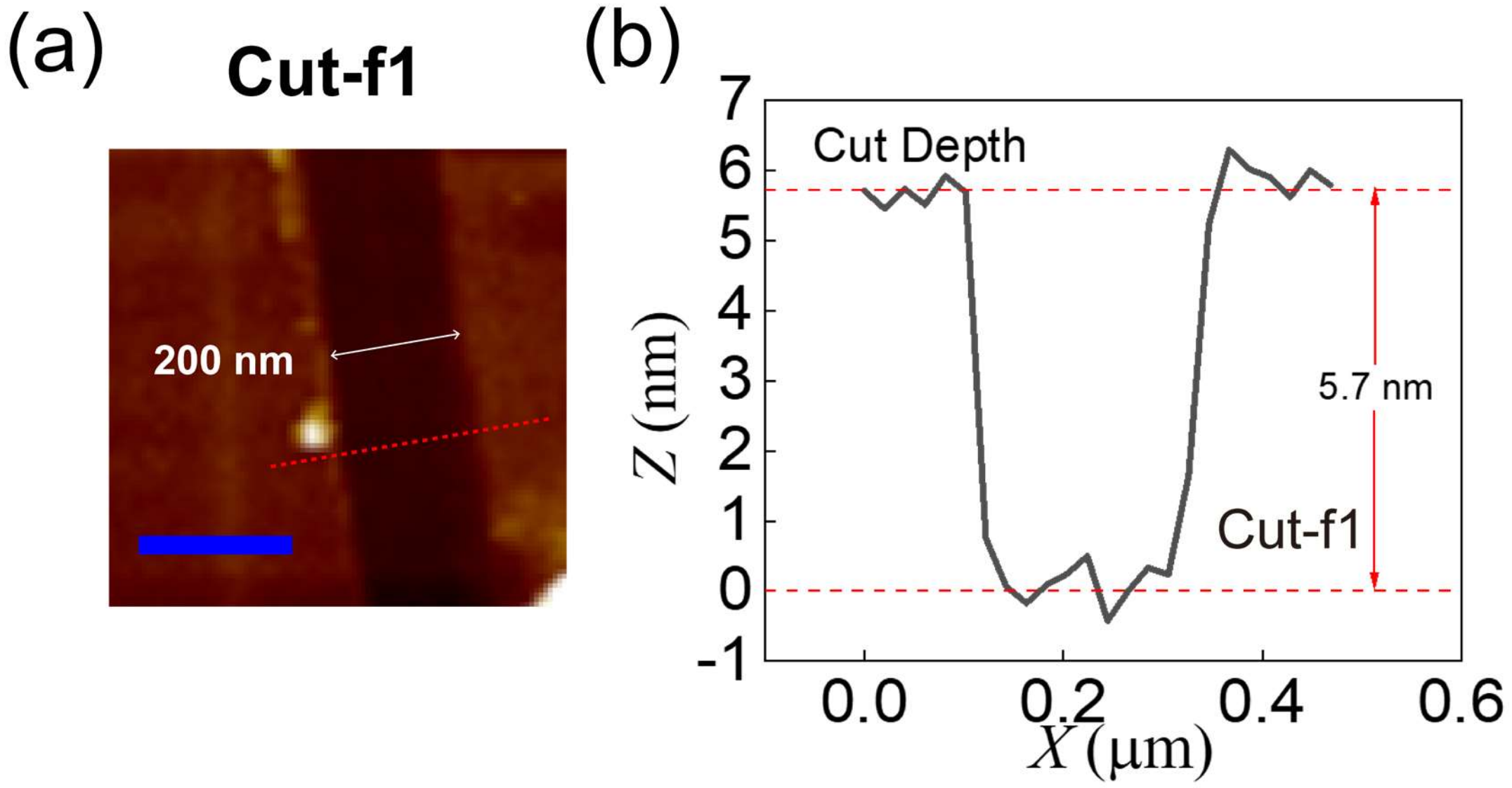

FIG. S8. AFM characterizations on the nanomachined cut in a MBT flake f1 with a thickness of 5.4 nm (4 SL). (a) AFM topographic image of the cut fabricated using our AFM-based nanomachining method. The interior of the trenches is clean and free of significant debris. The scale bar represents 250 nm. (b) Cross-sectional AFM line profile (*Z*-*X* curve) taken along the red dashed line in (a). The cut depth (5.7 nm) is slightly larger than the flake thickness (5.4 nm), confirming that the flake is completely cut through down to the substrate.

TABLE S1. Summary of experimental quantization results at 2 K compared with theoretical ideal values for the MBT devices with cuts. To quantitatively evaluate this consistency, we define the accuracy as $A = 1 - \frac{|R_{experimental} - R_{theoretical}|}{h/e^2}$.

| | | Experimental value ($h/e^2$) | Ideal value ($h/e^2$) | Accuracy |
|---|---|---|---|---|
| **Hall resistance** | device s2, B = 9 T | -0.990 | -1 | 99.0% |
| | device s2, B = -9 T | 0.990 | 1 | 99.0% |
| **longitudinal resistance** | device s2, B = 9 T | 0.001 | 0 | 99.9% |
| | device s2, B = -9 T | 0.001 | 0 | 99.9% |
| **three-terminal resistance** | device s2, B = 9 T | $<1 \times 10^{-4}$ | 0 | >99.99% |
| | device s2, B = -9 T | 0.984 | 1 | 98.4% |
| **two-terminal resistance** | device s2, B = 9 T | 0.987 | 1 | 98.7% |
| | device s2, B = -9 T | 0.987 | 1 | 98.7% |
| | device s3, B = 9 T | 0.980 | 1 | 98.0% |
| | device s3, B = -9 T | 0.980 | 1 | 98.0% |

TABLE S2. Summary of cut dimensions and transport data for the MBT devices with cuts. Near-quantized transport is retained in the measured devices over the investigated range of cut widths.

| Device Notation | Cut Notation | Trench Width (nm) | Trench Length ($\mu$m) | Longitudinal Resistance ($h/e^2$) | Two-terminal Resistance ($h/e^2$) |
|---|---|---|---|---|---|
| s2 | Cut-s2-A | 290 | 11.1 | 0.001 | 0.987 |
| | Cut-s2-B | 290 | 10.1 | | |
| s3 | Cut-s3-A | 10 | 0.7 | — | 0.980 |
| | Cut-s3-B | 120 | 10.8 | | |
| s4 | Cut-s4-A | 280 | 8.6 | 0.001 | — |
| | Cut-s4-B | 450 | 7.1 | | |